\documentclass[aps,prl,twocolumn,showpacs,amsmath,amssymb,superscriptaddress]{revtex4-1}

\usepackage{graphicx}
\usepackage{subfigure}
\usepackage{grffile}
\usepackage{epstopdf}
\usepackage{bm}
\usepackage{mhchem}
\usepackage{color}
\usepackage{hyperref}

\begin{document}

\title{Optical properties of monoclinic \ce{HfO2} studied by
  first-principles local density approximation + ${\bf U}$ approach}

\author{Jinping Li}
\email{lijinping@hit.edu.cn; jinping@yukawa.kyoto-u.ac.jp}
\affiliation{Center for Composite Materials, Harbin Institute of
  Technology, Harbin 150080, China}
\affiliation{Yukawa Institute for Theoretical Physics, Kyoto
  University, Kyoto, 606-8502, Japan}

\author{Jiecai Han}
\affiliation{Center for Composite Materials, Harbin Institute of
  Technology, Harbin 150080, China}

\author{Songhe Meng}
\affiliation{Center for Composite Materials, Harbin Institute of
  Technology, Harbin 150080, China}

\author{Hantao Lu}
\affiliation{Yukawa Institute for Theoretical Physics, Kyoto
  University, Kyoto, 606-8502, Japan}
\affiliation{Center for Interdisciplinary Studies $\&$ Key Laboratory
  for Magnetism and magnetic Materials of the MoE, Lanzhou University,
  Lanzhou 730000, China}

\author{Takami Tohyama}
\affiliation{Yukawa Institute for Theoretical Physics, Kyoto
  University, Kyoto, 606-8502, Japan}

\date{\today}

\begin{abstract}

  The band structures and optical properties of monoclinic \ce{HfO2}
  are investigated by the local density approximation (LDA)+$U$
  approach. With the on-site Coulomb interaction being introduced to
  $5d$ orbitals of Hf atom and $2p$ orbitals of O atom, the
  experimental band gap is reproduced. The imaginary part of the
  complex dielectric function shows a small shoulder at the edge of
  the band gap, coinciding with the experiments. This intrinsic
  property of crystallized monoclinic \ce{HfO2}, which is absent in
  both the tetragonal phase and cubic phase, can be understood as a
  consequence of the reconstruction of the electronic states near the
  band edge following the adjustment of the crystal structure. The
  existence of a similar shoulder-like-structure in the monoclinic
  phase of \ce{ZrO2} is predicted.

\end{abstract}



\maketitle

%\section{Introduction}

Hafnium dioxide (\ce{HfO2}) is widely studied both experimentally and
theoretically due to its excellent dielectric properties, wide band
gap, and high melting point, {\em etc.}~\cite{Weir:1999,Cao:1998}. It
has been widely used in optical and protective coatings, capacitors,
and phase shifting masks as one of the most promising high dielectric
constant materials~\cite{Gilo:1999, Yamamoto:2002, He:2007}.

There are three polymorphs of \ce{HfO2} existing at atmospheric
pressure~\cite{Terki:2008}: the monoclinic, the tetragonal and the
cubic fluorite, denoted as m-, t-, and c-\ce{HfO2}, respectively. It
has been known that the optical properties of thermally annealed
samples depend on their preparation process and therefore the
resulting structural details. In the thin films of m-\ce{HfO2} grown
on amorphous silica substrates, a small shoulder at the threshold of
the absorption spectra has been detected~\cite{Aarik:2004}. Following
experimental investigations have ruled out possible defect-related
origins~\cite{Park:2008}, and have noticed that the spectral weight of
the shoulder increases with the crystallite size in the
films~\cite{Kamala:2010}. On the other hand, no shoulder structure has
been observed in both the tetragonal phase and cubic phase. Therefore,
it is essential to investigate the electronic structure of \ce{HfO2}
and clarify the difference of optical properties in different
structural phases from the first-principles calculations.

The band structures of m-\ce{HfO2} have been calculated within the
framework of local density approximation (LDA) and generalized
gradient approximation (GGA), with further including spin-orbit
interactions~\cite{Garcia:2005}. However, the resulting band gap is
about $3.98\,\text{eV}$, much smaller than the experimental value,
$5.7\,\text{eV}$~\cite{He:2007}. The GW approximation, which can
address the electron correlations to the some extent, gives the gap
value as $5.78\,\text{eV}$, very close to the experimental
one~\cite{Jiang:2010}. Nevertheless, the GW method demands
considerable numerical resources. Another technique to include
correlation effect with less computational efforts is the so-called
LDA+$U$ or GGA+$U$ approach, where $U$ is the on-site Coulomb
interaction~\cite{Anisimov:1997}. Compared with LDA (GGA), the LDA
(GGA) + $U$ approach can produce qualitative improvements, e.g., see
Refs.~\cite{Anisimov:1997, Sun:2008, Zhang:2011}. A slightly extension
of the approach, i.e., the LDA +$U^f$ +$U^p$ and GGA +$U^d$ +$U^p$,
where the superscripts, $f$, $d$, and $p$, represent orbitals, has
been employed in the studies of \ce{CeO2} and c-\ce{HfO2},
respectively. And enhanced descriptions of the electronic structures
have been obtained~\cite{Plata:2012, Li:2013}.

%In court, it is reasonable to extend its applications to m-\ce{HfO2}.

In this paper, we use the LDA+$U$ scheme formulated by Loschen {\em{et
    al.}}~ \cite{Loschen:2007} to investigate the electronic
structures and optical properties of m-\ce{HfO2}. The on-site Coulomb
interactions of $5d$ orbitals on Hf atom ($U^d$) and of $2p$ orbitals
on O atom ($U^p$) are determined so as to reproduce the experimental
value of band gap. We find that the imaginary part of the (average)
dielectric function exhibits a shoulder structure at the edge of the
band gap, the existence of which is actually robust against the
perturbation with respect to $U$ values. A comparison study on
t-\ce{HfO2} and c-\ce{HfO2} confirms its absence in these phases. We
point out that the presence and absence of the shoulder can be
attributed to the difference of electronic structures near the edge of
the valence and conduction bands.

%\section{Computational Methodology}

In our calculation, the density functional theory simulations are
performed by using the LDA with CA-PZ functional and the LDA+$U$
approach as implemented in the CASTEP code (Cambridge Sequential Total
Energy Package)~\cite{Segall:2002}. For Hf and O atoms, the ionic
cores are characterized by plane-wave ultrasoft pseudopotentials. The
$5d^2$ and $6s^2$ electrons in Hf, $2s^2$ and $2p^4$ in O, are
explicitly treated as valence electrons. The plane-wave cut off energy
is $380\,\text{eV}$. The Brillouin-zone integration is performed over
the $24\times24\times24$ grid sizes using the Monkhorst-Pack method
for structure optimization. This set of parameters assure the total
energy convergence of $5.0\times10^{-6}\,\text{eV/atom}$, the maximum
force of $0.01\,\text{eV/\AA}$ , the maximum stress of
$0.02\,\text{GPa}$ and the maximum displacement of
$5.0\times10^{-4}\,\text{\AA}$. After optimizing the geometry
structure, we calculate the electronic structures and optical
properties of \ce{HfO2}. More numerical details can be found
elsewhere~\cite{Li:2013}.

%\section{Results and Discussion}

The space group of m-\ce{HfO2} is P21/c and the local symmetry is
C2h-5. The experimental values of the lattice constants $a$, $b$, $c$,
and the angle $\beta$ are following: $a=0.5117\,\text{nm}$,
$b=0.5175\,\text{nm}$, $c=0.5291\,\text{nm}$, and
$\beta=99.2^{\circ}$~\cite{Wang:1992}. The LDA calculation of the
perfect bulk m-\ce{HfO2} is performed to determine the optimized
parameters in order to check the applicability and accuracy of the
ultrasoft pseudopotential. The results, $a=0.5225\,\text{nm}$,
$b=0.5349\,\text{nm}$, $c=0.5365\,\text{nm}$, and
$\beta=99.5^{\circ}$, are in good agreement with
experiments~\cite{Wang:1992} and other theoretical
values~\cite{Zhao:2002, Caravaca:2005, Tan:2012}. However, the value
of the band gap $E_g$ is around $3.24\,\text{eV}$, much smaller than
the experimental value ($\sim 5.7\,\text{eV}$). This is due to the
fact that the density functional theory usually undervalues the energy
of $5d$ orbitals of Hf atom, lowering the bottom level of conduction
bands.

\begin{figure}
\subfigure[]{\includegraphics[width=0.22\textwidth]{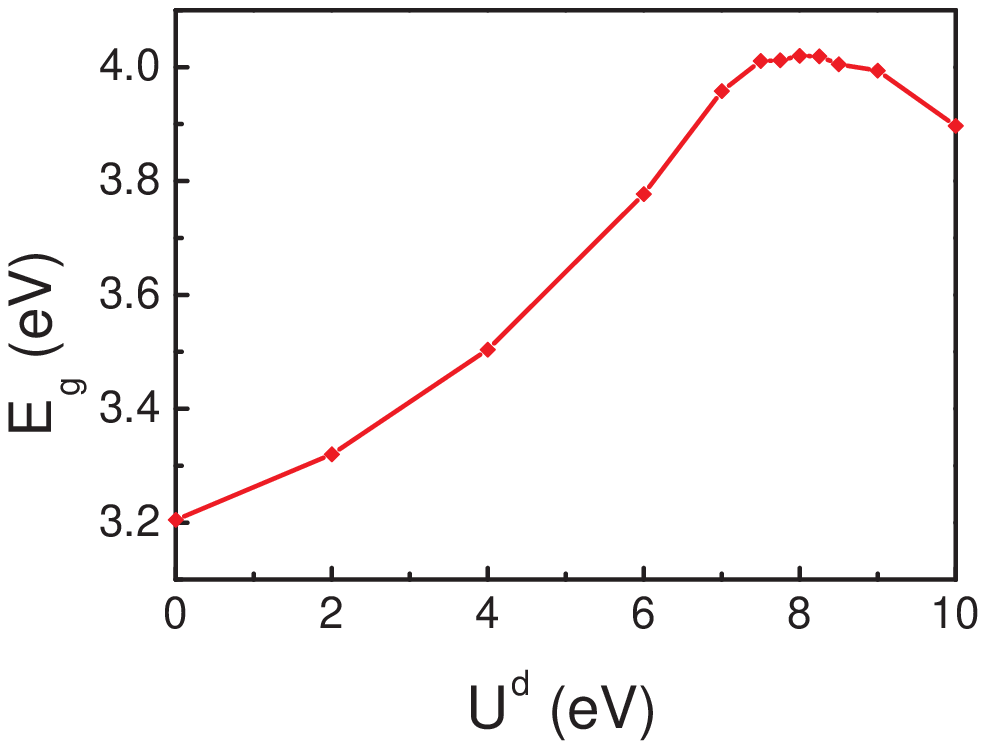}
\label{fig1a}}
\subfigure[]{\includegraphics[width=0.215\textwidth]{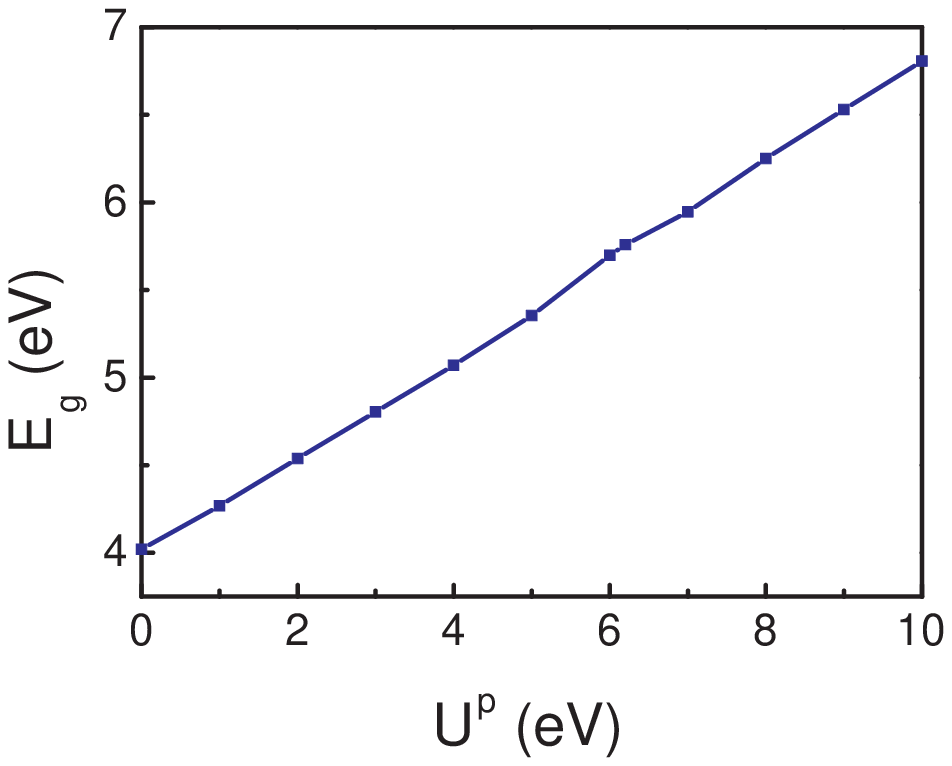}
\label{fig1b}}
\caption{(Color online) The band gap $E_g$ as a function of (a) $U^d$
  and (b) $U^p$.}
\label{fig1}
\end{figure}

In order to reproduce the band gap, we first introduce $U^d$ for $5d$
orbitals of Hf atom. Using the experimental lattice parameters as
initial values, we optimize geometry structure and calculate the band
structure and density of state (DOS) of m-\ce{HfO2}. The band gap
$E_g$ obtained from the band structure is shown in Fig.~\ref{fig1a} as
a function of $U^d$. It can be seen that $E_g$ firstly increases, and
then drops with the increase of $U^d$, showing a maximum value ($\sim
4.02\,\text{eV}$) at $U^d=8.0\,\text{eV}$, where the lattice
parameters of the optimized structure are $a=0.5382\,\text{nm}$,
$b=0.5370\,\text{nm}$, $c=0.5474\,\text{nm}$, and
$\beta=99.7^{\circ}$. The maximum $E_g$ value is still smaller than
the experimental one. The saturation of $E_g$ with $U^d$ may be
related to the approach of $5d$ states toward $6s$ and $5p$ states,
though microscopic mechanism is not yet fully understood. Next, we
introduce $U^p$ for $2p$ orbital of O atom, while keeping $U^d$ fixed
at $8.0\,\text{eV}$. Different from Fig.~\ref{fig1a}, the results in
Fig.~\ref{fig1b} shows a monotonic increase in $E_g$ as a function of
$U^p$. When $U^d=8.0\,\text{eV}$ and $U^p=6.0\,\text{eV}$, the
calculated band gap of m-\ce{HfO2} is $5.70\,\text{eV}$, well
coinciding with the experiment. The lattice parameters of the
optimized structure are $a=0.5386\,\text{nm}$, $b=0.5331\,\text{nm}$,
$c=0.5492\,\text{nm}$, and $\beta=99.6^{\circ}$.

\begin{figure}
\subfigure[]{\includegraphics[width=0.22\textwidth]{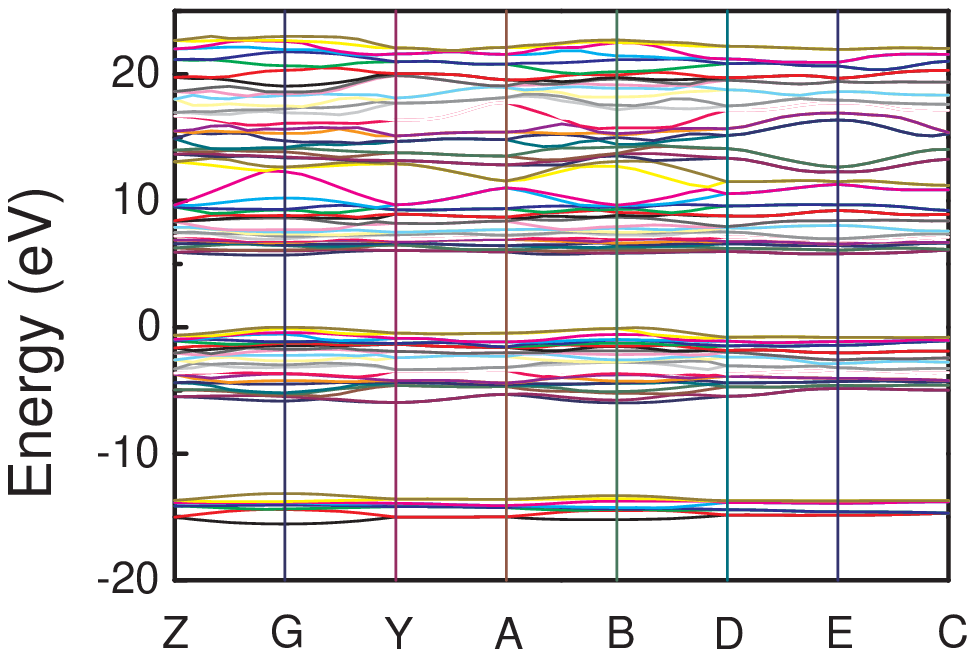}
\label{fig2a}}
\subfigure[]{\includegraphics[width=0.22\textwidth]{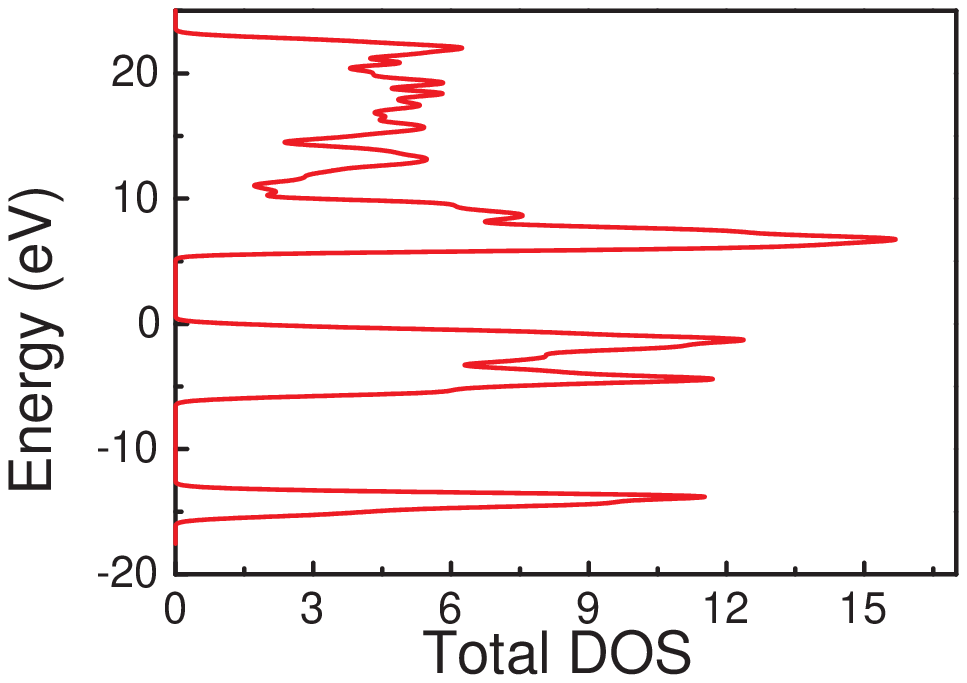}
\label{fig2b}}
\\
\subfigure[]{\includegraphics[width=0.22\textwidth]{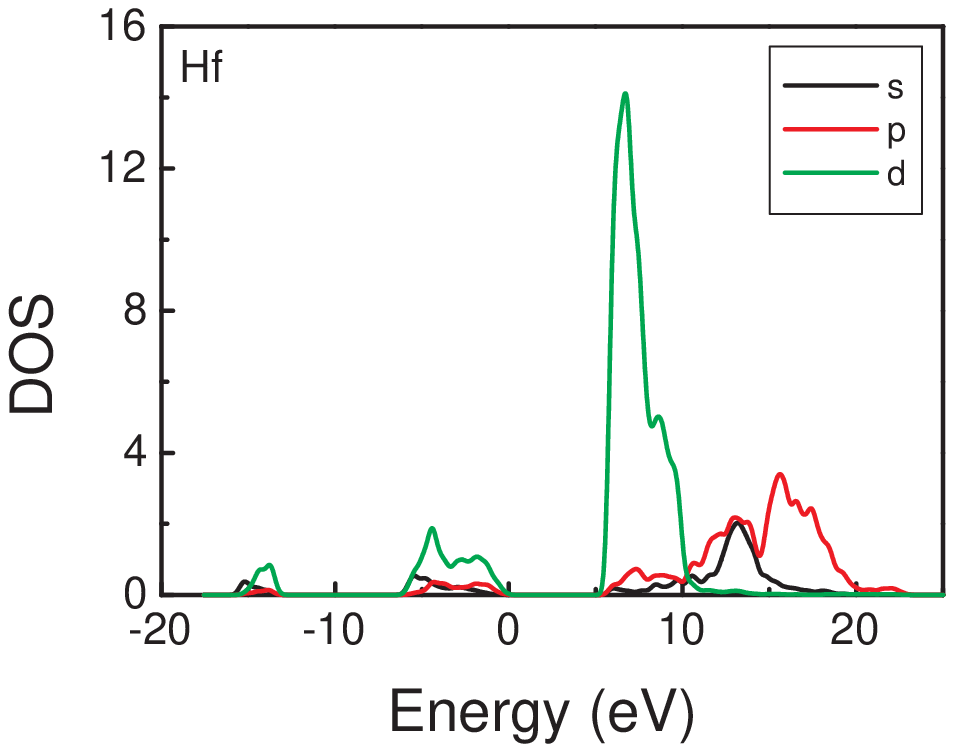}
\label{fig2c}}
\subfigure[]{\includegraphics[width=0.22\textwidth]{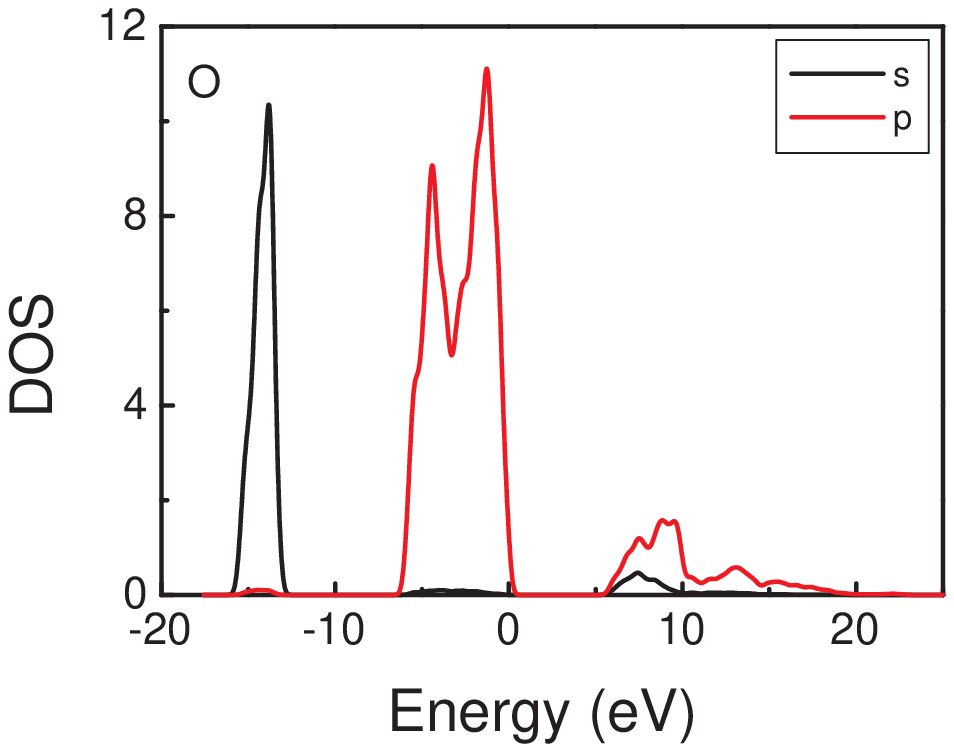}
\label{fig2d}}
\caption{(Color online) The band structure and density of states (DOS)
  of m-\ce{HfO2} obtained by LDA+$U^d$+$U^p$ ($U^d=8.0\,\text{eV}$,
  $U^p=6.0\,\text{eV}$). (a) Band structure. The total DOS, the
  partial DOS of Hf and O atoms are shown in (b), (c) and (d),
  respectively.}
\label{fig2}
\end{figure}

By adopting the optimal $U$ values as $U^d=8.0\,\text{eV}$,
$U^p=6.0\,\text{eV}$, we perform the LDA+$U$ calculation. The band
dispersion is presented in Fig.~\ref{fig2a}. The bottom of the
conduction band is located at the G point. Since the bottom is lifted
to higher energy by introducing $U^d$, accompanied with the
reconstruction of the conduction band, the DOS around $4.5\,\text{eV}$
and $6.9\,\text{eV}$, which is separated in LDA without $U$ (not
shown), has merged into one sharp structure at $6.8\,\text{eV}$
(Fig.~\ref{fig2b})~\cite{Li:2013}. According to the partial DOS of Hf
and O atoms in Figs.~\ref{fig2c} and \ref{fig2d}, we can see that just
above and below the gap, the conduction band is predominantly
constructed by Hf $5d$ states, while the valence band by O $2p$
states. Therefore, the low-lying optical excitations across the gap is
mainly composed by the interband transitions from the O $2p$ to the Hf
$5d$ orbitals.

\begin{figure}
\hspace*{-0cm}
%\hskip -1.6cm
\subfigure[]{\includegraphics[width=0.25\textwidth,angle=-90]{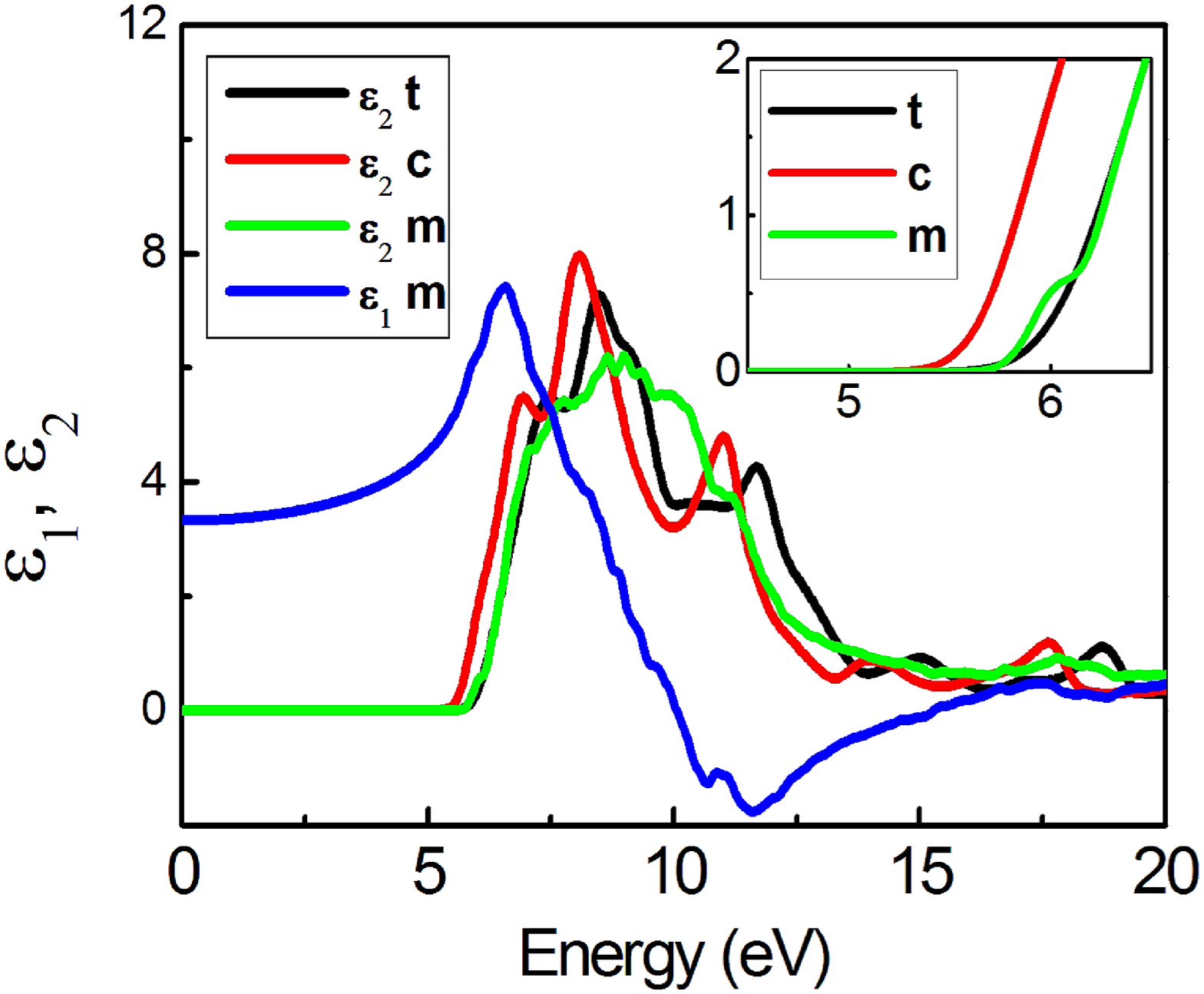}
\label{fig3a}}
%\hspace*{5cm}
%\hskip -1.4cm
\subfigure[]{\includegraphics[width=0.25\textwidth,angle=-90]{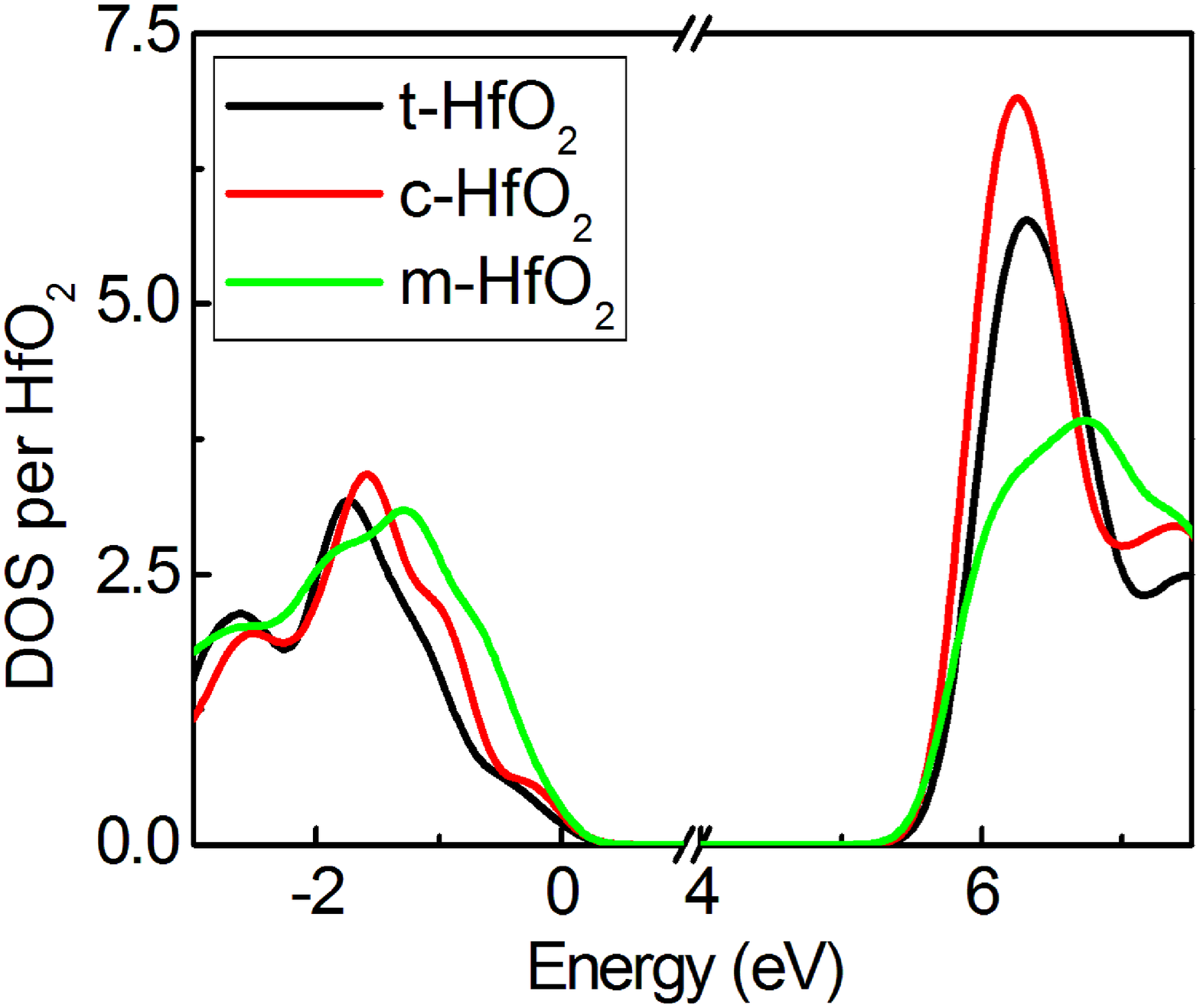}
\label{fig3b}}
\caption{(Color online) Comparison of (a) dielectric functions and (b)
  total DOS of m-\ce{HfO2} with t-\ce{HfO2} and c-\ce{HfO2} obtained
  by LDA+$U^d$+$U^p$. Inset in (a): Zoom-in view of the imaginary
  part of the dielectric function, $\epsilon_2$ near the gap edge.}
\label{fig3}
\end{figure}

Figure~\ref{fig3a} shows the dielectric function of m-\ce{HfO2}, with
comparison to those of t-\ce{HfO} and c-\ce{HfO}. The real part,
$\epsilon_1$, exhibits a maximum at $6.57\,\text{eV}$.  The calculated
static dielectric constant is $3.32$, coinciding with the experimental
value~\cite{He:2007, Koike:2006}. The imaginary part, $\epsilon_2$,
shows a maximum at $9.0\,\text{eV}$. The maximum, around $6.2$, is
very close to the experimental observation $6.0$~\cite{Park:2008},
while the value obtained by LDA without $U$ is around $8.7$. Other
optical properties, like optical conductivity, can be computed from
the complex dielectric function~\cite{Wang:1992}. We also obtain the
refractive coefficient $n=1.82$, close to the experimental value
$1.93$~\cite{He:2007}.

The calculated $\epsilon_2$ of t-\ce{HfO2} and c-\ce{HfO2} are
presented in Fig.~\ref{fig3a} for comparison. The values of $U^d$ and
$U^p$ are determined so as to reproduce the experimental gap
values. Here, the resulting $U^d$=$8.25\,\text{eV}$
($8.25\,\text{eV}$) and $U^p$=$6.3\,\text{eV}$ ($6.25\,\text{eV}$) for
t-\ce{HfO2} (c-\ce{HfO2}~\cite{Li:2013}). We find that $\epsilon_2$
exhibits similar spectral distributions between t-\ce{HfO2} and
c-\ce{HfO2}, with only a small shift along the energy direction ($\sim
1\,\text{eV}$). By contrast, the global spectral distribution of
$\epsilon_2$ in m-\ce{HfO2} is quite different from them. More
interestingly, as shown in the inset of Fig.~\ref{fig3a}, a small
shoulder structure emerges at the edge of the band gap ($\sim
6.0\,\text{eV}$) in m-\ce{HfO2}, which, on the other hand, cannot be
found either in t-\ce{HfO2} or in c-\ce{HfO2}. The exclusive presence
of the shoulder structure is consistent with experimental
observations~\cite{Aarik:2004, Park:2008, Kamala:2010}.

In order to understand the origin of the shoulder, we compare the
total DOS of the valence and conduction bands in m-\ce{HfO2} with
those in the other two phases. As shown in Fig.~\ref{fig3b}, the DOS
in t-\ce{HfO2} and c-\ce{HfO2} are similar to each other, while there
is significant difference in m-\ce{HfO2}. In particular, at the edge
of the valence band, the DOS in m-\ce{HfO2} smoothly increases but
there is a step feature in t-\ce{HfO2} and c-\ce{HfO2} at the
edge. Furthermore, the conduction band in m-\ce{HfO2} reveals a broad
feature of DOS around $6.5\,\text{eV}$ in contrast to t-\ce{HfO2} and
c-\ce{HfO2}, in which a peak appears near $6.0\,\text{eV}$. Since
$\epsilon_2$ is given by the excitation from the valence band to the
conduction band across the band gap, the difference of DOS between
m-\ce{HfO2} and t-(c-)\ce{HfO2} is indicative of the presence of the
shoulder in $\epsilon_2$ only for m-\ce{HfO2}. A more detailed
microscopic origin of the shoulder, such as the assignment of the
momentum and band index dominating the absorption at the shoulder,
remains to be resolved in the future. We note that the shoulder
structure also appears at the edge of the band gap in the standard LDA
calculation without $U$ (not shown). This means that the
reconstruction of the electronic states due to the monoclinic
structure is crucial for the shoulder structure, as indicated by the
previous experimental investigations~\cite{Aarik:2004}.

\begin{figure}
\begin{center}
\includegraphics[width=0.22\textwidth,angle=270]{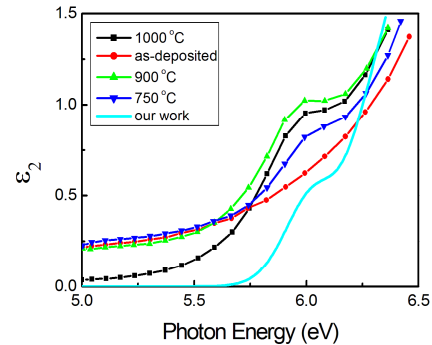}
\caption{(Color online) The imaginary part of the dielectric function
  for m-\ce{HfO2} calculated by the LDA+$U^d$+$U^p$ (the solid (or sky
  blue) line), compared with the experimental data on various
  conditions (taken from Ref.~\cite{Park:2008}).}
\label{fig4}
\end{center}
\end{figure}

The experimental studies for \ce{HfO2} have shown that thermally
annealed films~\cite{Park:2008}, crystallite films~\cite{Plata:2012},
and thin films grown on amorphous silica substrates~\cite{Aarik:2004}
reveal a shoulder-like feature in $\epsilon_2$. The experimental data
showing the annealing effect taken from Ref.~\cite{Park:2008} are
plotted together with the present theoretical result in
Fig.~\ref{fig4}. We find that the calculated small shoulder appears at
almost the same energy ($\sim 6.0\,\text{eV}$) as experimental
shoulder for the annealed samples. This means that the present
LDA+$U^d$+$U^p$ approach can reproduce an essential feature of the
electronic structures in crystalline m-\ce{HfO2}. We note that the
magnitude of $\epsilon_2$ is different from the experimental ones. The
difference may partly come from the fact that the samples of
m-\ce{HfO2} reported in Ref.~\cite{Park:2008} mix with small amount of
orthorhombic and tetragonal \ce{HfO2}.

Finally, it is worth to mention that we have additionally carried out
the calculation on \ce{ZrO2}, using the same strategy. In many
aspects, \ce{ZrO2} resembles its twin oxide, \ce{HfO2}, though the
electron correlations are generally believed to be weaker. The
shoulder-like feature in the dielectric function is also found in the
monoclinic phase of \ce{ZrO2}, while absent in the other two phases
(tetragonal and cubic), similar to other theoretical results obtained
from full-relativistic calculation~\cite{Garcia:2006}. This seems not
consistent with experiments, since no shoulder-like structure has been
reported ~\cite{Garcia:2006, Shin:2011, Lin:2003}. Here we would like
to point out that in order to prepare pristine crystallized \ce{ZrO2}
with monoclinic structure, the temperature of thermal annealing should
reach over $1,000\,^{\circ}\text{C}$~\cite{Tang:2005}, which means
that up to now, for the \ce{ZrO2} samples used in the optical
measurements, the monoclinic component may not be dominant. Based on
our results, we predict the emergence of the shoulder-like structure
in the imaginary part of the dielectric function with the monoclinic
\ce{ZrO2} (m-\ce{ZrO2}) being prevalent in the mixture of the three
phases.

%\section{Conclusions}

In summary, the on-site Coulomb interactions for the $5d$ orbital of
Hf atom ($U^d$) and the $2p$ orbital of O atom ($U^p$) are introduced
into the first-principles LDA band calculation of m-\ce{HfO2}. With
the optimal values of $U^d =8.0\,\text{eV}$, $U^p =6.0\,\text{eV}$,
the experimental band gap is reproduced. A shoulder-like structure at
the edge of the band gap in the imaginary part of the dielectric
function is obtained, which is consistent with the experiments. The
presence of the shoulder in m-\ce{HfO2} and its absence in t-\ce{HfO2}
ad c-\ce{HfO2} indicate the impact of the crystal structure on the
electronic bands and optical properties. The existence of a similar
shoulder-like structure in m-\ce{ZrO2} is predicted.

%\begin{acknowledgments}

This work was supported by the Foundation
for Innovative Research Groups of the National Natural Science
Foundation of China (Grant No. 11121061), the Strategic Programs for
Innovative Research (SPIRE), the Computational Materials Science
Initiative (CMSI) and the Yukawa International Program for
Quark-Hadron Sciences at YITP, Kyoto University. T.T. acknowledges
support by the Grantin-Aid for Scientic Research (Grant No. 22340097)
from MEXT.

%\end{acknowledgments}

\end{document}